\documentclass[%
 reprint,
 amsmath,
 amssymb,
 prb,
 floatfix
]{revtex4-2}

\usepackage{graphicx}
\usepackage{dcolumn}
\usepackage{bm}
\usepackage{xcolor}
\usepackage{subcaption}
\usepackage{multirow}

\newtheorem{definition}{Definition}[section]
\usepackage{hyperref}

\begin{document}


\title{Sparse expansions of multicomponent oxide configuration energy using coherency \& redundancy}

\author{Luis Barroso-Luque}
\email{lbluque@berkeley.edu}
\affiliation{Department of Materials Science and Engineering, University of California Berkeley, Berkeley CA, USA}
\author{Julia H. Yang}
\affiliation{Department of Materials Science and Engineering, University of California Berkeley, Berkeley CA, USA}

\author{Gerbrand Ceder}
\email{gceder@berkeley.edu}
\affiliation{Department of Materials Science and Engineering, University of California Berkeley, Berkeley CA, USA}
\affiliation{Materials Sciences Division, Lawrence Berkeley National Laboratory, Berkeley CA, USA}

\date{\today}

\begin{abstract}
Compressed sensing has become a widely accepted paradigm to construct high dimensional cluster expansion models used for statistical mechanical studies of atomic configuration in complex multicomponent crystalline materials. However, strict sampling requirements necessary to obtain minimal coherence measurements for compressed sensing to guarantee accurate estimation of model parameters are difficult and in some cases impossible to satisfy due to the inability of physical systems to access certain configurations. Nevertheless, the dependence of energy on atomic configuration can still be adequately learned without these strict requirements by using compressed sensing by way of coherent measurements using redundant function sets known as frames. We develop a particular frame constructed from the union of all occupancy-based cluster expansion basis sets. We illustrate how using this highly redundant frame yields sparse expansions of the configuration energy of complex oxide materials that are competitive and often surpass the prediction accuracy and sparsity of models obtained from standard cluster expansions.
\end{abstract}

\pacs{Valid PACS appear here}
\maketitle


\section{Introduction} \label{sec:intro}

The cluster expansion method (CEM)\cite{sanchezGeneralizedClusterDescription1984, sanchezClusterExpansionsConfigurational1993, vandervenFirstPrinciplesStatisticalMechanics2018} has become a standard tool in computational thermodynamics of multicomponent crystalline systems as an effective and efficient means to compute functions of atomic configuration. The CEM is used to represent a coarse-grained lattice model of materials properties as a function of the possible atomic configurations, i.e., the occupations of crystallographic sites. The most common use of the CEM involves the expansion of the formation energy of a particular multicomponent material system. A cluster expansion is fitted using formation energies for a set of representative structures calculated from first principles electronic structure methods, most often calculated with density functional theory (DFT). Well-converged cluster expansions fitted to DFT energies have been shown to give very accurate predictions of formation energy to within a few meV/atom.\cite{vandewalleAutomatingFirstprinciplesPhase2002a, sanchezFoundationsPracticalImplementations2017, vandervenFirstPrinciplesStatisticalMechanics2018} A converged cluster expansion can subsequently be used in Monte Carlo simulations to compute thermodynamic properties, such as configuration entropy,\cite{otto_relative_2013, liu_monte_2021} phase diagrams,\cite{drichards_design_2016} ionic percolation,\cite{urban_configurational_2014, ouyang_effect_2020} and short range order.\cite{clement_short-range_2018, kostiuchenko_short-range_2020}

The CEM involves the use of a discrete harmonic expansion in terms of basis functions---known by practitioners as correlation functions---that act on groups of symmetrically related clusters of crystallographic sites. The correlation functions used in the CEM are constructed by taking symmetry adapted averages of all possible \textit{$N$-fold} products of site basis functions for all $N$ sites in a crystal structure. Site basis functions operate over a domain associated with each site in a structure, which we refer to as a site space. The site functions must span the site space associated with each given site. In the CEM the site functions constitute a basis for the space of functions acting over the allowed occupations at a site, such that the number of site basis functions must be equal to the number of allowed species (i.e. the size of the site space). Additionally, in order for the resulting product basis to have a cluster-like framework, where basis functions operate on different clusters of sites, each set of site basis functions must include a constant function $\phi_0 \equiv 1$ as one of the functions used to span the corresponding occupations for each site.\cite{vandervenFirstPrinciplesStatisticalMechanics2018}

Estimating the correlation function coefficients, known as effective cluster interactions (ECI), for a specific expansion is most often done using linear regression.\cite{connollyDensityfunctionalTheoryApplied1983, vandervenFirstPrinciplesStatisticalMechanics2018, sanchezFoundationsPracticalImplementations2017} The values of a predefined finite set of correlation functions---chosen within a spatial cutoff range and maximum cluster size---are computed for each training structure to form a truncated correlation matrix $\Pi_S = A\Pi$, where $\Pi$ is the full system (infinite for bulk systems) correlation matrix, and $A$ is a \emph{sensing} matrix that selects the set of correlation functions to be considered in the fit, and also selects the values of those correlation functions for each structure included in the set of training structures. A column of $\Pi_S$ is made up of the values of an included correlation function evaluated for each of the training structures, and are correspondingly feature vectors. A row of $\Pi_S$ is called a correlation vector and its entries correspond to each of the included correlation functions evaluated for a particular structure in the training set.

Several linear regression algorithms for estimating ECI have been proposed and benchmarked in literature.\cite{vandewalleAutomatingFirstprinciplesPhase2002a, blum_using_2005, hart_evolutionary_2005-1,  muellerBayesianApproachCluster2009, nelsonCompressiveSensingParadigm2013, leongRobustClusterExpansion2019a} Although most of the proposed algorithms can be used in either overdetermined and underdetermined linear systems, the formal analysis of solutions and mathematical guarantees on the accuracy of coefficients can be very different and are for the most part carried out for each type of linear system separately.\cite{calafiore_elghaoui_2014, hastie_elements_2001} In over-determined systems more training structures are included than the number of correlation functions used in the fit. The motivation for using an over-determined system comes from several studies showing decreasing cross validation errors with increasing number of training structures.\cite{leongRobustClusterExpansion2019a, angqvist_icet_2019} In the case of underdetermined systems, more correlation functions are included in the model than structures used when fitting; which is usually motivated for the CEM as a way to avoid a nearsighted pre-selection of correlation functions. In the current work, we focus on the case of using underdetermined linear systems using $\ell_1$ regularization as a convex relaxation of the $\ell_0$ pseudo-norm, in order to fit sparse expansions of configuration energy. This approach can be formally cast and analyzed within the framework of Compressed Sensing (CS).\cite{candesIntroductionCompressiveSampling2008, candesCompressedSensingCoherent2011a}

For relatively small dimensional systems, such as binary or ternary alloys, that allow accurate fits using only a small number of short range clusters of low degree (number of sites in cluster) such that the number of total correlation functions is manageable, practitioners have the option to choose between using an overdetermined system or an underdetermined system approach. However, for more complex high dimensional systems there will often be no such choice. Since the number of expansion terms that need to be considered for an acceptable fit quickly becomes too large, computing DFT energies for enough structures to obtain an overdetermined system becomes untenable. The reason for this is twofold. First of all, the number of terms in a truncated expansion grows polynomially with the number of allowed species at each site. Additionally, complex physical interactions may require longer range and/or higher degree clusters. For example, these situations frequently occur in the study of high entropy materials which has recently gained much attention in the design of metallic alloys,\cite{george_high-entropy_2019} and ceramic materials.\cite{lun_cation-disordered_2021} Although theoretically the CEM is well suited for the study of high entropy materials, the large high dimensional configuration spaces involved render an \textit{over-determined} system approach prohibitive. For such scenarios, there is no choice but to work with underdetermined systems only, and hope that \textit{betting on sparsity}\cite{hastie_elements_2001} applies favorably to high dimensional cluster expansions for the increasing number of materials being studied that require fits using severely underdetermined systems.

Fortunately, classical compressed sensing\cite{candesIntroductionCompressiveSampling2008} in the CEM has already been demonstrated to work well with some metallic alloy systems.\cite{nelsonCompressiveSensingParadigm2013, nelsonClusterExpansionMade2013} Here we go beyond previous work and apply the theory of compressed sensing with coherency and redundancy to the realm of the CEM, and demonstrate how it applies in much larger configuration spaces for ionic systems with multiple substructures and oxidation state assignment. In Section \ref{sec:cs} we introduce the main concepts of compressed sensing with coherence and redundancy and restate the main signal recovery error bound\cite{candesCompressedSensingCoherent2011a} adapted to the domain of cluster expansions. In Section \ref{sec:ce} we lay out the details of cluster expansions using site indicator basis functions. This is subsequently used as a building block in the construction of a redundant function set, which we call the \textit{generalized Potts frame}, which we describe in Section \ref{sec:potts}. Finally, we demonstrate how the generalized Potts frame yields sparse and accurate expansions by comparing resulting fits of three different fluorinated lithium-transition metal oxide materials with standard cluster expansions using site indicator functions and orthogonal trigonometric/sinusoid basis functions.\cite{vandewalleMulticomponentMultisublatticeAlloys2009}

\section{Compressed Sensing with Coherency \& Redundancy} \label{sec:cs}

The notion of classical compressed sensing (CS) that has been previously shown to yield accurate cluster expansions of metallic alloys,\cite{nelsonCompressiveSensingParadigm2013, nelsonClusterExpansionMade2013} relies strictly on the concept of \textit{incoherence}\cite{candesIntroductionCompressiveSampling2008} in order to guarantee accurate recovery of the underlying coefficients/ECI. The need for incoherent measurements---which in the CEM correspond to correlation function values evaluated for selected training structures---is clear when the goal is to accurately recover the exact coefficients in the expansion of a function. This requirement can be made intuitive by thinking of \textit{coherence} as a measurement of how similar correlation function samples (i.e., columns of $\Pi_S$) are to each other. High \textit{incoherence} (low \textit{coherence}), means that the set of correlation function samples used are more uniformly spread out in their span (they are closer to mutually orthogonal), and as a result the portion of the energy represented by each, can be easily distinguished.
In the limit of zero coherence, the correlation matrix is orthogonal, and in principle the portion of the energy for each correlation function can be identified exactly. For the cases with high coherence, correlation function samples will be much more closely correlated and it is no longer possible to distinguish which correlation function a specific portion of the energy comes from.

Formally, the \textit{coherence} of a measurement matrix $M$---which in the CEM corresponds the truncated correlation matrix $\Pi_S$---is defined as,\cite{davenportIntroductionCompressedSensing2012}
\begin{equation}
    \mu(M) = \underset{i < j}{\max}\frac{|\langle M_i, M_j\rangle|}{||M_i||_2||M_j||_2}
\end{equation}
where $M_i$ and $M_j$ are columns of matrix $M$. The normalized definition given above bounds the coherence of a measurement matrix to values between zero and one, $0\le \mu(M) \le 1$.

In the case of the CEM, the measurement matrix $M = A\Pi$ is made up of the values of a relatively small set of truncated correlation vectors for each of the training structures for which DFT energies have been calculated, and the truncated correlation vectors only include functions associated with clusters within a physical predefined cutoff radius and cluster size.

The guarantees of classical compressed sensing depend on the coherence of the measurement matrix $M$ being as close to zero as possible in order to maximize the probability of accurate reconstruction of coefficients/ECI. As suggested above, incoherent measurements improve the chances of accurate recovery of coefficients/ECI, specifically by requiring a lower number of necessary training structures for a given level of accuracy.\cite{candesIntroductionCompressiveSampling2008} Furthermore, it is necessary for $M$ to satisfy the \textit{restricted isometry property} (RIP) with a small isometry constant  $\delta$.\cite{candesIntroductionCompressiveSampling2008} In broad terms, the RIP is a condition which ensures that most of the possible sparse solutions for the linear system lie outside the nullspace of the measurement matrix $M$.\cite{davenportIntroductionCompressedSensing2012}

It has been shown that matrices made up of random measurements, such as those composed of Gaussian vectors on the unit hypersphere, satisfy the RIP with overwhelming probability and lead to high levels of incoherence.\cite{candesIntroductionCompressiveSampling2008} Random measurements have been previously reported to lead to effective training structure selection and resulting accurate underdetermined fits of CE's for binary metallic alloys.\cite{nelsonCompressiveSensingParadigm2013, nelsonClusterExpansionMade2013} Nevertheless, selecting training structures such that the resultant measurement matrix has high incoherence and ideally satisfies the RIP with a small constant, becomes difficult and in some cases almost impossible for materials systems with more complex physics. This can be generally understood as a result of physically imposed sampling constraints. Usually the vast majority of all possible configurations will have high energies that are complicated if not impossible to compute with first principles methods. For example, systems with configurations that undergo large structural relaxations can no longer be mapped back to the fixed structure underlying the CE. As another example, certain cluster occupations in ionic systems are difficult to access when very high electrostatic repulsion exists. Finally, charge neutrality constraints in heterovalent ionic systems restricts the possible configurations that can be sampled. These phenomena complicate and in many cases prevent the possibility of obtaining appropriate correlation matrices with minimal coherence required for classical CS recovery, even with previously proposed methods, such as the aforementioned use of uniformly random vectors over the hypersphere.

\begin{figure*}
    \includegraphics[width=0.7\textwidth]{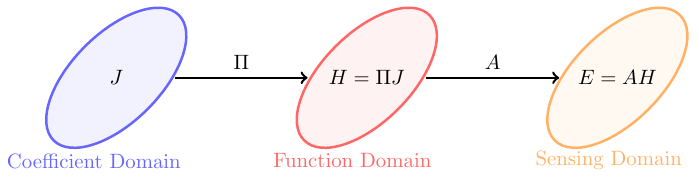}
    \caption{Schematic of different domains involved in compressed sensing. CS with redundancy seeks to recover function $H$ in the function domain in the center. Adapted from Candes \emph{et al}.\cite{candesCompressedSensingCoherent2011a}}
    \label{fig:domains}
\end{figure*}

Fortunately, there is a variant of compressed sensing that has been shown to give accurate reconstructions of sparse or compressible signals with only a small set of \textit{coherent} measurements by including redundancy in these measurements.\cite{candesCompressedSensingCoherent2011a} We will show how this concept can be used for learning sparse expansions of functions of configuration in multicomponent crystalline materials. The essence of CS with coherent and redundant measurements is not to recover the model coefficients as accurately as possible, but rather to recover an approximation for the actual function as accurately as possible by way of a redundant representation and possibly highly coherent measurements.\cite{candesCompressedSensingCoherent2011a} For systems constructed using measurements with high coherence and a redundant representation, solutions will always be highly degenerate, but again the focus is on the accuracy of predictions obtained using the recovered expansion and not the exact values of the fitted coefficients. However, we will still seek only sparse solutions such that the large degeneracy of solutions becomes manageable and the resulting expansion can be used efficiently for subsequent predictions and simulations such as in Monte Carlo studies. Figure \ref{fig:domains} shows a schematic of the pertinent mathematical objects, the corresponding domains and their relationships as it pertains to CS. For our purposes we seek an accurate representation in the function domain made up from a small number of coefficients but without much regard to whether or not the obtained coefficients correspond to those of any underlying \textit{exact} expansion.

In what follows we only give a brief overview of CS with coherent and redundant measurements adapted to our purposes. Further details and accompanying proofs can be found in the original publication.\cite{candesCompressedSensingCoherent2011a}

The goal of CS, is to reconstruct a sparse representation or approximation of a function by solving the following optimization problem.
\begin{equation}
    \hat{H} = \underset{\widetilde{H}}{\text{argmin}}||\Pi^*\widetilde{H}||_1 \text{  subject to } ||A\widetilde{H} - E ||_2 \le \varepsilon
    \label{eq:cs-problem}
\end{equation}

where $H$ is the function sought---in our case a function of atomic configuration. $\Pi$ is a linear operator mapping coefficients $J$ to the function $H$ as depicted in Figure \ref{fig:domains}. In the CEM, $\Pi$ is the matrix of all correlation basis functions. $A$ is a sensing matrix used in selecting the training data; and $E$ is the measured value of the function $H$ for the training configurations. The measurements are of the form $E = AH + z$, where $z$ represents an additive noise with some upper bound $||z||_2\le\varepsilon$.

The main theorem of CS with coherency and redundancy  gives the following error bounds for an $s$-sparse reconstruction of $H$,
\begin{equation}
    ||\hat{H} - H||_2 \le C_0\varepsilon + C_1 \frac{||\Pi^*H - (\Pi^*H)_s||_1}{\sqrt{s}}
    \label{eq:cs-theorem}
\end{equation}
where $\hat{H}$ is the $s$-sparse approximation to $H$. $(\Pi^*H)_s$ denotes a vector with the largest $s$ nonzero coefficients of $H$ and zeros elsewhere. $C_0$ and $C_1$ are constants that depend only on the sensing matrix $A$. The theorem holds if the measurement matrix satisfies a restricted isometry property (D-RIP) adapted to the union of the span of all sets of $s$ columns of $\Pi$ with isometry constant $\delta_{2d} < 0.08$.\cite{candesCompressedSensingCoherent2011a} The D-RIP prevents possible solutions from being highly distorted by the measurement matrix, and similarly to the standard RIP, also prevents them from falling in its null space.

The theorem from Equation \ref{eq:cs-theorem} essentially says that the solution to the optimization problem in Equation \ref{eq:cs-problem} gives accurate reconstructions of the function $H$, when the coefficients $\Pi^*H$ are sparse and/or decay rapidly---i.e., $H$ is \textit{compressible}. These results suggest that if functions of configuration for multicomponent materials are indeed \textit{compressible}, meaning they can be represented with only a small set of functions, then accurate reconstructions can be obtained with coherent and redundant measurements. Although a rigorous proof along with bounds on the compressibility of functions of configuration---which to the best of our knowledge has not been reported in literature---would be of immense value, it is beyond the scope of the present work. We do however provide numerical indication that such functions are indeed favorably compressible. This result is not unanticipated considering established knowledge regarding the physics of locality, the success of atomic potentials with small numbers of multiple body terms, and the general tenet of parsimony in physics. This translates to our intuition that such expansions of atomic configuration should have low degree and small associated cluster diameters, and as such we should be able to represent them using a small number of terms. The crux is finding an optimal subspace spanned by a small set of functions that allows an accurate and sparse representation.

\section{The Cluster Expansion with Site Occupancy Basis Functions} \label{sec:ce}

Before introducing the generalized Potts frame, we give a brief exposition of the process of constructing a basis in the CEM using a particular choice of site basis functions. Although the original CEM was developed using \textit{orthonormal} site basis functions constructed from polynomials,\cite{sanchezGeneralizedClusterDescription1984, sanchezClusterExpansionsConfigurational1993} any set of functions can be used as long as they properly span the function space over the corresponding site spaces. Here we will focus on one commonly used basis set composed of site \textit{occupancy} functions,\cite{zhangClusterExpansionsThermodynamics2016} which we refer to as site \textit{indicator} functions. A site \textit{indicator} function simply \textit{indicates} whether a given species occupies a site or not,
\begin{equation}
    \mathbf{1}_{\sigma_j}(\sigma_i) =
              \begin{cases} 
                  1 & \text{if } \sigma_i = \sigma_j\\
                  0 & \text{otherwise}
              \end{cases}
\end{equation}

Since the set of basis functions must include the constant function for the CEM to yield basis functions that follow a cluster framework, the constant function needs to replace an indicator function for one of the species at each site. In other words, one of the total $n$ allowed species at each site will not have an associated indicator function. \cite{zhangClusterExpansionsThermodynamics2016} The sets of site basis functions $\{\phi_j; j=0,\ldots, n-1\}$ for the CEM in terms of site indicator functions are given by,
\begin{equation}
    \phi_j(\sigma_i) = \begin{cases} 
                  1 & \text{if } j=0\\
                  \mathbf{1}_{\sigma_j}(\sigma_i) &\text{if } j = 1,\ldots,n - 1
              \end{cases}
    \label{eq:site-basis}
\end{equation}

The correlation functions are then constructed from symmetry adapted averages of the following $N$-fold products of site basis functions for each site in a structure,
\begin{equation}
    \Phi_{\bm{\alpha}}(\bm{\sigma}) = \prod_{i=1}^N\phi_{\alpha_i}(\sigma_i),
    \label{eq:product-basis}
\end{equation}
where $\bm{\alpha}$ is a multi-index used to specify the particular site function, $\alpha_i = 0,\ldots, n_i-1$, for each site in the structure. Since we can ignore all sites where the constant site basis function $\phi_0 \equiv 1$ is used, Equation \ref{eq:product-basis}, can be written in terms of site indicator functions as,
\begin{equation}
    \Phi_{\bm{\alpha}}(\bm{\sigma}) = \prod_{i \text{ s.t. } \alpha_i\neq 0}\mathbf{1}_{\sigma_{\alpha_i}}(\sigma_i),
    \label{eq:indicator-product-basis}
\end{equation}

where we notice that by construction, the functions in Equation \ref{eq:indicator-product-basis} will indicate whether a specific occupancy of a given cluster represented by the nonzero elements of the multi-index $\bm{\alpha}$ is present in a structure. We can more briefly write Equation \ref{eq:indicator-product-basis} as a \textit{cluster} indicator function,
\begin{align}
    \Phi_{\bm{\alpha}}(\bm{\sigma}) &= \mathbf{1}_{\bm{\alpha}}(\bm{\sigma}) \nonumber\\
    &= \begin{cases} 
              1 & \text{if cluster } \bm{\alpha} \text{ is in } \bm{\sigma} \\
              0 & \text{otherwise}
          \end{cases}
    \label{eq:cluster-indicator}
\end{align}

Furthermore, since the final correlation functions are constructed from averages of functions given by Equation \ref{eq:cluster-indicator} over symmetrically equivalent clusters, the correlation functions represent (up to multiplicity constants) the \textit{concentration} of a specific occupancy of clusters for the different crystallographic orbits,
\begin{equation}
    \langle \Phi(\bm{\sigma})\rangle_{\beta} = \frac{1}{N(\beta)}\sum_{\bm{\alpha} \in \beta} \mathbf{1}_{\bm{\alpha}}(\bm{\sigma})
    \label{eq:corr-functions}
\end{equation}
where $\beta$ represents an orbit of labeled clusters (i.e. with each site labeled with a particular indicator function), and $N(\beta)$ is the total number of labeled clusters in the orbit $\beta$. The sum is carried over all labeled clusters $\bm{\alpha}$ that are part of the orbit $\beta$.

We make two notable observations regarding CEM correlation functions with site indicator basis functions. The first is that the basis sets in Equation \ref{eq:site-basis} are not orthogonal and as a result the resulting correlation functions in Equation \ref{eq:corr-functions} are not orthogonal either. This lack of orthogonality can further complicate constructing highly incoherent measurement matrices compared to using orthogonal/orthonormal basis sets.

The second observation is concerned with the set of clusters/orbits that are selected by the correlation functions in Equation \ref{eq:corr-functions}. The set of correlation functions in Equation \ref{eq:corr-functions} indeed represents a basis for the function space over all possible configurations $\bm{\sigma}$ and thus the set is linearly independent.\cite{zhangClusterExpansionsThermodynamics2016} As a result however, the included functions do not give the concentrations of all possible occupied clusters. Namely, the correlation functions in Equation \ref{eq:corr-functions} never include any occupied clusters that involve the species that do not have an associated indicator function in the corresponding site basis functions. Additionally, the number of clusters not indicated for in the CEM basis grows quickly with the number of components. Namely the number grows as $O\left(n^{N_{\bm{\alpha}}} - (n - 1)^{N_{\bm{\alpha}}}\right)$, where $n$ is the number of allowed species at a site and $N_{\bm{\alpha}}$ is the number of sites in a cluster.

Notwithstanding these observations, cluster expansions using site indicator functions formally constitute a basis for function spaces over crystalline configurations, and have been successfully used in the study of configurational thermodynamics for some time.\cite{zhangClusterExpansionsThermodynamics2016, drichards_design_2016, ouyang_effect_2020} However, as alluded to before, the lack of orthogonality complicates obtaining incoherent measurements to maximize accurate ECI estimation from classical CS. Additionally, the choice of species left out in site basis sets is mathematically meaningless, but can often lead to precarious interpretations of fitted ECI. This begs the question of whether we can do away with using CEM basis sets and seeking maximally incoherent measurements, yet still obtain suitably sparse and accurate expansions by instead relying on redundancy by including functions labeled for all possible occupations over the included clusters.

\section{The Generalized Potts Frame as a Redundant Representation} \label{sec:potts}

We introduce a specific redundant set of functions that spans the same function space as correlation basis sets used in the CEM. The redundant set of functions can be obtained simply by including cluster indicator functions of the form in Equation \ref{eq:indicator-product-basis} for all possible occupations. More formally we build all the $N$-fold product functions from redundant site function sets that include $\phi_0\equiv 1$ and site indicator functions for \textit{all} allowed species at each site, such that for each site with $n$ allowed species, we associate a redundant set of functions given by,
\begin{equation}
    \phi_j(\sigma_i) = \begin{cases} 
                  1 & \text{if } j=0\\
                  \mathbf{1}_{\sigma_j}(\sigma_i) &\text{if } j = 1, \ldots, n
              \end{cases}
    \label{eq:site-frame}
\end{equation}

This results in $n + 1$ total functions for a space of dimension $n$, where the redundancy is the trivial linear relation,
\begin{equation}
    \phi_0(\sigma) - \sum_{i=1}^n\mathbf{1}_{\sigma_i}(\sigma) = 0
    \label{eq:triv-linear}
\end{equation}

\begin{figure*}
    \includegraphics[width=0.8\textwidth]{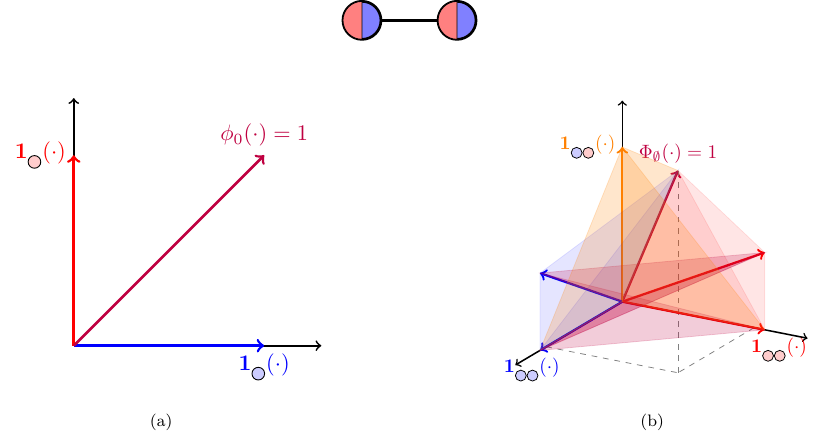}
    \caption{Function space representations over the configurations of a single binary site and a symmetric binary diatomic molecule. (a) Function space over a single binary site space. The two different choices for site bases to construct a standard CE are colored red and blue, and each set also includes the purple $\phi_0 \equiv 1$ function. (b) Function space over symmetrically distinct configurations of the molecule. A CE basis includes either the blue colored or the red colored functions. The generalized Potts frame includes all colored functions (blue/red/yellow). All functions sets also include the magenta colored constant function. The D-RIP for a 2-sparse representation in this case is adapted to the union of all colored planes in (b).}
    \label{fig:subspaces}
\end{figure*}

Carrying out the same operation of taking all possible $N$-fold products of functions from each redundant set and taking symmetry adapted averages over crystallographic orbits results in the set of symmetrized product functions which give \textit{concentrations} for any possible cluster occupancy. This resulting set is clearly highly redundant since the total number of functions obtained are of order $O\left((n + 1)^N\right)$ and the dimension of the function space is of order $O(n^N)$. In other words the combinatorics\cite{hartAlgorithmGeneratingDerivative2008} for identifying symmetrically equivalent clusters involve $n$ possible \textit{labels} of basis functions for each site for all clusters in distinct orbits compared with the $n-1$ \textit{labels} involved in a standard CEM basis. Nonetheless, the resulting set of functions spans the same function space as a standard CEM basis and therefore formally the set constitutes a \textit{frame}.\cite{christensen_frames_2008, waldron_introduction_2018} We provide a derivation for a set of frame bounds in Appendix \ref{sec:bounds}, but make no effort to optimize them. The resulting frame evidently has a strong connection to the well known Potts model.\cite{potts_generalized_1952} Indeed it is a (normalized) generalization in both spatial extent and interaction size of the original nearest neighbor pair Potts model.\cite{wu_potts_1982} Hence we refer to the proposed frame as the \textit{generalized Potts frame}.

The generalized Potts frame can also be seen as the union of every possible CEM correlation basis of site indicator functions. All possible CEM correlation bases can be generated by cycling over the species that is not \textit{indicated} for in its corresponding site basis and building the corresponding CEM correlation basis for every possible combination of site basis sets. We make a special note that apart from the standard CEM basis sets, this includes correlations of \textit{mixed} products where symmetrically equivalent sites in the underlying random structure can have distinct basis sets, i.e. a different subset of species with associated indicator functions. Symbolically, the generalized Potts frame is obtained after taking the corresponding symmetry adapted averages for the product functions in the following set,
\begin{equation}
    \bigcup_{\sigma}\left\{\Phi_{\bm{\alpha}};\;\forall \bm{\alpha}\; \text{ s.t. } \bm{\alpha}\text{ is not in }\bm{\sigma} \right\}
\end{equation}

Though the generalized Potts frame is highly redundant, the motivation is that by introducing more expansion functions than strictly necessary and using an appropriate algorithm---such as for solving Equation \ref{eq:cs-problem}---can yield both accurate and sparse representations of functions of crystalline configurations without the need of maximally incoherent measurements. An intuition for this can be formed by picturing the union of all subspaces spanned by size $s$ subsets of functions in the generalized Potts frame. If the function sought lies on any of these subspaces or close enough, then the function can be accurately represented by only $s$ terms.

As a simple illustration consider a symmetric binary diatomic system. Figure \ref{fig:subspaces}a shows the corresponding site spaces and site indicator functions along with the constant function $\phi_0$. Additionally, the resulting symmetrized product bases are also shown in Figure \ref{fig:subspaces}b. The basis functions in both cases are colored to represent the possible CE bases. The union of all basis functions from these CE bases corresponds to the generalized Potts frame for this simple diatomic system. Again we highlight also the inclusion of products of \textit{mixed} site basis sets, which corresponds to the orange vector built from the product of a red and blue basis functions associated with each site respectively. In this case the function space over configurations of the binary diatomic  molecule has three dimensions, however when a function lies close to or on any of the highlighted planes that  function can be well approximated using only two terms.

\section{Applications to Multi-component Oxides} \label{sec:apps}

\begin{figure*}
    \begin{subfigure}{0.3\textwidth}
        \includegraphics[width=\textwidth]{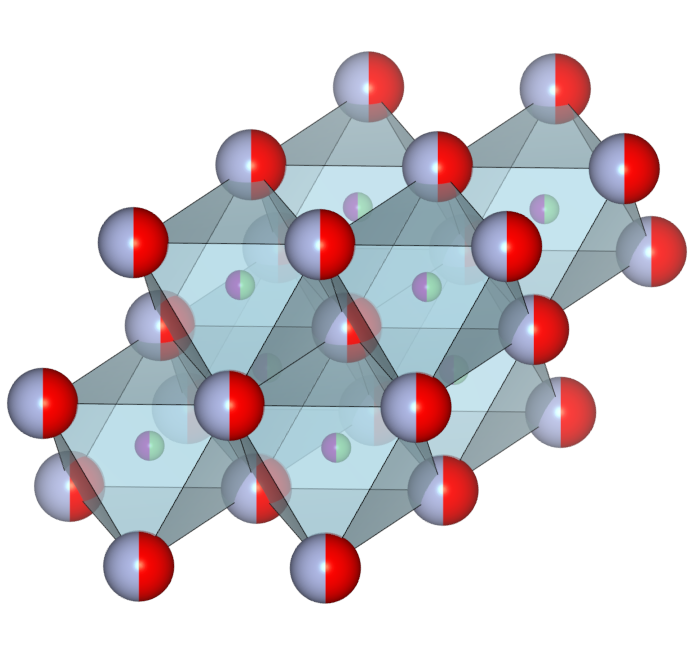}
        \caption{2-2 LiMnOF binary-binary rocksalt}
    \end{subfigure}
    \begin{subfigure}{0.3\textwidth}
        \includegraphics[width=\textwidth]{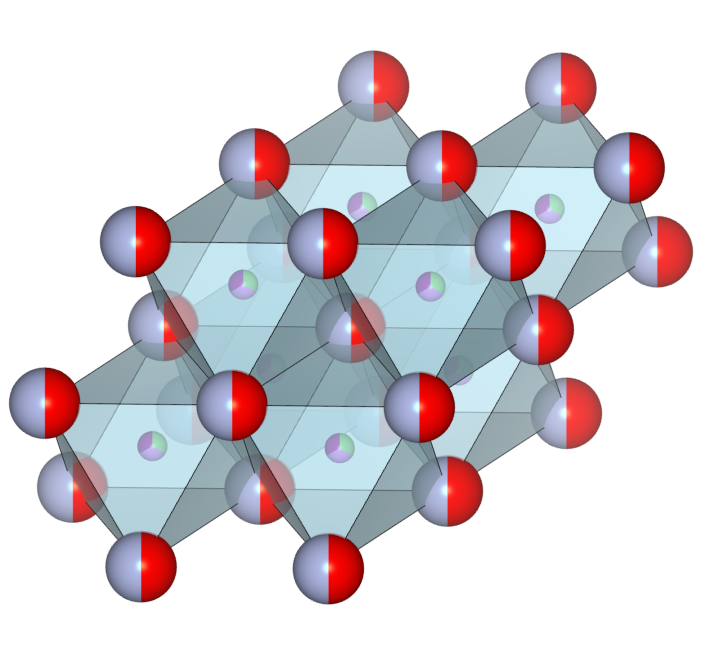}
        \caption{3-2 LiMnTiOF ternary-binary rocksalt}
    \end{subfigure}
    \begin{subfigure}{0.3\textwidth}
        \includegraphics[width=\textwidth]{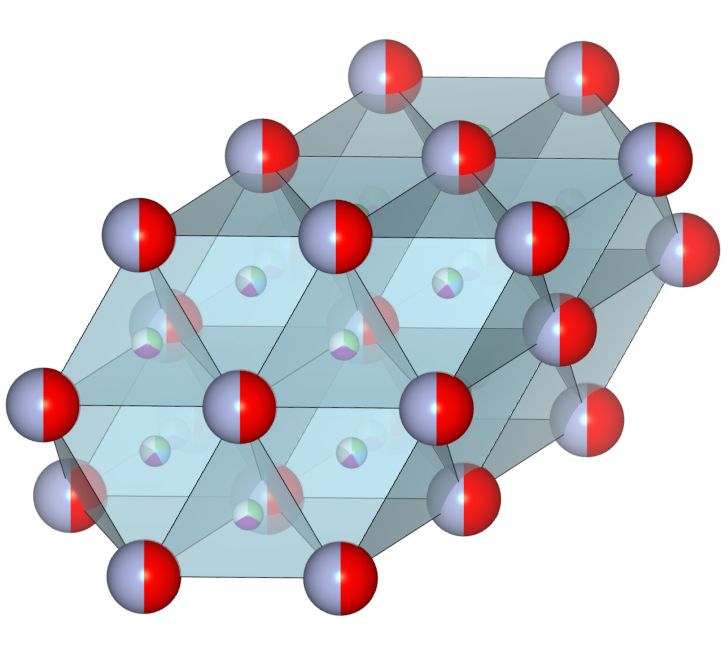}
        \caption{5-3-2 LiMnOF quinary-ternary-binary spinel-like}
    \end{subfigure}
    \caption{(a) System 2-2: Li-Mn-O-F rocksalt system with binary (Li+/Mn3+) cation sites and binary (O2-/F-) anion sites. A primitive cell with two sites is used as the formula unit for normalization. (b) System 3-2: Li-Ti-Mn-O-F rocksalt system with ternary (Li+/Mn3+/Ti4+) cation sites and binary (O2-/F-) anion sites. A primitive cell with two sites is used as the formula unit for normalization. (c) System 5-3-2: Li-Mn-O-F spinel-like system with quinary (Li/Mn2+/Mn3+/Mn4+/vacancy) octahedral cation sites, ternary (Li/Mn2+/vacancy) tetrahedral cation sites, binary (O2-/F-) anion sites. A primitive cell with four sites is used as the formula unit for normalization.}
    \label{fig:structures}
\end{figure*}

We demonstrate the performance of the generalized Potts frame by comparing fits for three fluorinated lithium-transition metal oxide systems with fits obtained using a site indicator based CE (with site basis functions given in Equation \ref{eq:site-basis}), and a CE using orthogonal sinusoid site basis functions.\cite{vandewalleMulticomponentMultisublatticeAlloys2009} These material systems are of interest as novel cathode materials made with abundant elements,\cite{clement_cation-disordered_2020} however their specific applications are not particularly relevant for the purpose of this work. The configuration spaces for the materials considered increase in both size and complexity (larger number of allowed species and number of symmetrically distinct sites) as shown in Figure \ref{fig:structures}.

\begin{table*}
    \begin{tabular}{c c c c c}
         \hline\hline
         System &Expansion Type &Training Set Size &Test Set Size &Model Size \\ \hline
         \multirow{2}{*}{2-2} &CEM &112 &337 &121\\
         &Potts Frame &112 &337 &520 \\
         \multirow{2}{*}{3-2} &CEM &195 &456 &312\\
         &Potts Frame &195 &456 &1040 \\ 
         \multirow{2}{*}{5-3-2} &CEM &312 &56 &7030$^*$ \\
         &Potts Frame &312 &56 &23070$^*$ \\ \hline\hline
    \end{tabular}
    \caption{Regression model and training/test data size specifications for the three fluorinated lithium-transition metal oxide systems. $^*$Removal of correlation functions that remained constant for structures in the training set reduced the number of columns in the measurement matrices in 5-3-2 system to $4194$ and $17350$ for the indicator basis CEM based and Potts frame models respectively.}
    \label{tab:terms}
\end{table*}

The total number of expansion terms considered for each fit are listed as the model size in Table \ref{tab:terms}. The total number of terms are obtained by using the same cutoff radius for clusters with up to four sites. Functions that evaluate to the same value (remain constant) for the training structures used in the corresponding fit are subsequently removed from the final measurement matrix used for each fit. Particularly, removal of constant functions was only required for the 5-3-2 system, for which the total number of columns in the measurement matrix ended up being $4194$ and $17350$ for the indicator basis CEM based and Potts frame models respectively. Additionally, we include an electrostatic energy term\cite{drichards_design_2016, sekoClusterExpansionMulticomponent2014a} as an additional feature in every fit. Finally, for the fits using the Potts frame, we remove one cluster indicator function from each set associated with the same orbit; this is simply to do away with the trivial linear relation in Equation \ref{eq:triv-linear} applied to cluster concentrations. Doing so has a minimal effect in reducing redundancy, however we found this slightly improved efficiency for obtaining full rank measurement matrices. It is clear by construction that for the same spatial cutoffs, the number of terms in the generalized Potts frame far exceeds the number of terms in a standard cluster expansion, and this difference grows exponentially with the configuration space complexity. 

The number of training structures and test structures for each fit is also listed in Table \ref{tab:terms}. The structures were generated by using an initial set of structures from the Materials Project\cite{Jain2013}, and subsequently using this set and Monte Carlo to generate additional structures of up to 216 atoms. The formation energy for each training/test structure was computed with density functional theory using the Vienna ab initio simulation package\cite{kresse_efficiency_1996} using the projector augmented wave (PAW) method,\cite{kresse_ultrasoft_1999} with reciprocal space discretization of 25 k-points per \AA \ and a plane wave energy cutoff of 520 eV. We carry out spin-polarized calculations using the SCAN meta-GGA exchange correlation \cite{sun_strongly_2015} and pseudopotentials which include semicore states: Li\_sv, Mn\_pv, O, and F. Structures are converged to $10^{-6}$ eV in total energy and 0.01 eV/\AA \ on atomic forces.

A total of 50 different fits for each system are computed by selecting a random set of training structures that gives a full rank underdetermined system. The remaining structures are used as a test set. The regression problem solved which is known as \textit{Basis Pursuit Denoising} or the \textit{LASSO} is slightly different than the $\ell_1$-analysis problem given in Equation \ref{eq:cs-problem}. The corresponding optimization problem is,

\begin{equation}
    \bm{\hat{J}} = \underset{\bm{\widetilde{J}}}{\text{argmin}}\; ||\bm{\widetilde{J}}||_1 \text{  subject to  } ||A\Pi\bm{\widetilde{J}} - E||_2 \le \varepsilon
    \label{eq:bpursuit}
\end{equation}

The problem in Equation \ref{eq:bpursuit} is solved for each fit using the python package \texttt{scikit-learn}\cite{scikit-learn_2011} in a two-step process. First a 10-fold hyperparameter cross validation optimization search is done with the training set. Subsequently a finer hyperparameter search is done centered at the previously obtained value now optimizing for out-of-sample error with respect to the test set but training only with the training set data. From the resulting fits, those with sparsity (number of nonzero coefficients) above the third quartile of the set are considered outliers and removed from the results.

\begin{figure}
    \includegraphics[width=0.45\textwidth]{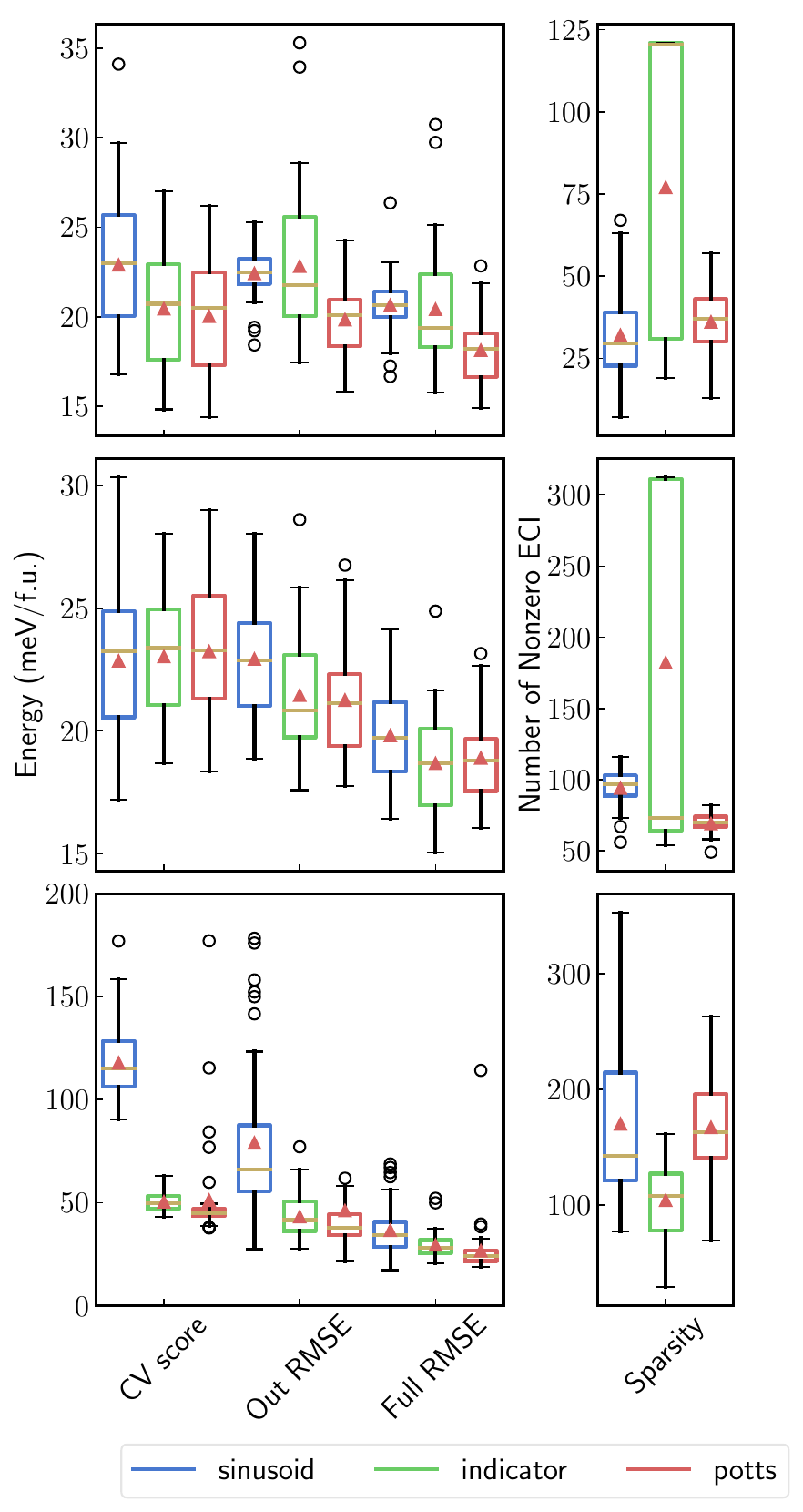}
    \caption{Fit metric statistics for the systems tested using standard CEM with sinusoid site basis, indicator site basis and generalized Potts frame. The plotted metrics include: cross validation RMSE (CV score), out of sample RMSE (Out RMSE), full data RMSE (Full RMSE) for both the training and test structures combined, and the number of nonzero ECI in the fits (sparsity). LiMnOF binary-binary with two sites per formula unit (top), LiMnTiOF ternary-binary with two sites per formula unit (middle), LiMnOF quinary-ternary-binary with four sites per formula unit (bottom).}
    \label{fig:fit-stats}
\end{figure}

Values for accuracy metrics and sparsity results for the fitted expansion obtained from the set of fits for each of the three expansion types and for each of the three materials systems are shown in boxplots in Figure \ref{fig:fit-stats}. The average prediction accuracy metrics given include cross validation root mean squared error (CV RMSE) for the initial cross validation hyperparameter search, the out of sample RMSE for the final fit (test structures only), and the full data RMSE for both the training and test structures combined. The average sparsity value for the resulting fits is also listed. 

Figure \ref{fig:fit-stats} shows boxplots depicting the resulting fit statistics in terms of cross validation, out of sample and full sample (both training and testing structures) root mean squared errors, along with the corresponding model sparsity values. The results depicted in Figure \ref{fig:fit-stats} show that for all systems, the expansions resulting from the Potts frame tend to have either a lower minimum, median and mean accuracy metrics, a lower spread in these accuracy metrics or both. In all cases the accuracy metrics are either very competitive (very close to standard cluster expansions), or better. As for the resulting model sparsity, although considerably more terms are included in the Potts frame fit, the resulting expansions have similar sparsity along with lower spread in sparsity values. These results demonstrate the applicability of using coherency and redundancy to obtain expansions that match and even exceed accuracy and sparsity of those obtained with standard cluster expansion basis sets.

\begin{table*}
    \begin{tabular}{cccccc}
         \hline\hline
         System &Expansion &CV RMSE &Out RMSE &Full RMSE &Sparsity \\\hline
         \multirow{3}{*}{2-2}
         &Sinusoid &34.10 &25.26 &26.36 &7 \\
         &Indicator &23.16 &19.37 &18.79 &19 \\
         &Potts Frame &23.82 &22.31 &21.5 &13 \\
         \multirow{3}{*}{3-2}
         &Sinusoid &24.13 &27.62 &24.14 &56 \\
         &Indicator &25.14 &21.69 &19.55 &54 \\
         &Potts Frame &22.26 &23.43 &20.93 &49 \\
         \multirow{3}{*}{5-3-2}
         &Sinusoid &131.76 &158.16 &68.80 &77 \\
         &Indicator &45.52 &49.99 &52.12 &29 \\
         &Potts Frame &44.58 &40.51 &39.65 &69 \\
         \hline\hline
    \end{tabular}
    \caption{Fitted model accuracy metrics and sparsity of sparsest models. Cross validation RMSE (CV RMSE), out of sample RMSE (out RMSE), and full dataset RMSE (full RMSE) in meV per formula unit (random structure primitive cell).}
    \label{tab:sparsest-results}
\end{table*}

The boxplots in Figure \ref{fig:fit-stats} provide a summary of the average values obtained for each type of fit, however they do not allow us to see which models have both high accuracy and high sparsity (low number of terms). Tables \ref{tab:sparsest-results} and \ref{tab:accurate-results} list the same fit metrics as Figure \ref{fig:fit-stats} but for the single sparsest (smallest number of nonzero coefficients) fit and the fit resulting in the highest accuracy in terms of the full data RMSE for each materials system.

\begin{table*}
    \begin{tabular}{cccccc}
         \hline\hline
         System &Expansion &CV RMSE &Out RMSE &Full RMSE &Sparsity \\
         \hline
         \multirow{3}{*}{2-2}
         &Sinusoid &25.68 &18.42 &16.66 &50 \\
         &Indicator &18.63 &17.87 &15.76 &121 \\
         &Potts Frame &21.42 &16.11 &14.89 &43 \\
         \multirow{3}{*}{3-2}
         &Sinusoid &23.10 &18.88 &16.44 &105 \\
         &Indicator &28.02 &17.89 &15.07 &310 \\
         &Potts Frame &23.09 &17.77 &16.06 &72 \\
         \multirow{3}{*}{5-3-2}
         &Sinusoid &104.39 &36.72 &17.24 &291 \\
         &Indicator &55.69 &27.45 &20.38 &147 \\
         &Potts Frame &115.42 &24.52 &18.98 &192 \\
         \hline\hline
    \end{tabular}
    \caption{Fitted model accuracy metrics and sparsity of most accurate models in terms of the root mean squared error on the whole dataset. Cross validation RMSE (CV RMSE), out of sample RMSE (out RMSE), and full dataset RMSE (full RMSE) in formula unit (random structure primitive cell).}
    \label{tab:accurate-results}
\end{table*}

From Table \ref{tab:sparsest-results}, we see that for the sparsest expansions obtained, the ones fitted using the Potts frame result in the lowest error metrics with only a few exceptions where the Potts frame fits are still of comparable accuracy. Furthermore, the expansions based on the Potts frame result in the lowest or second lowest sparsity for all three systems. The results for the most accurate models in terms of full dataset RMSE in Table \ref{tab:accurate-results} show that the Potts frame results in both sparse and accurate models. For the cases where one of the CEM based fits results in a lower full RMSE, the sparsity of that model is compromised and substantially worse than the corresponding Potts frame fit. This behavior can also be observed in the results in Figure \ref{fig:fit-stats}, where the box plot for sparsity obtained with the Potts frame has a smaller interquartile range than that of the CEM based fits (with the exception of the 5-3-2 system where it is only slightly larger than that of the indicator based cluster expansion).

\begin{figure}
    \includegraphics[width=0.5\textwidth]{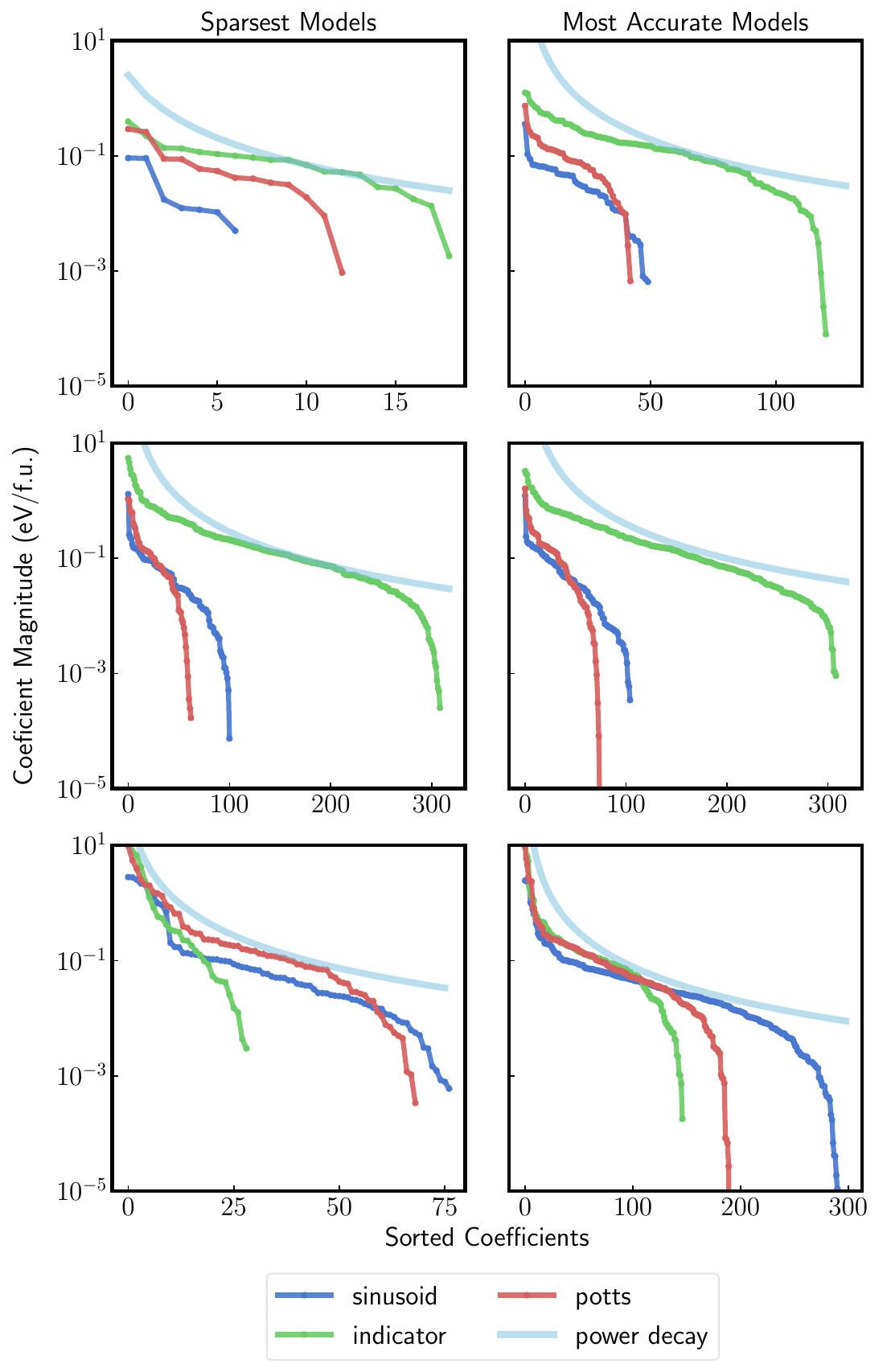}
    \caption{Sorted fitted coefficient magnitudes (multiplicity times ECI) for sparsest and most accurate model (Full RMSE). LiMnOF binary-binary (top), LiMnTiOF ternary-binary (middle), LiMnOF quinary-ternary-binary (bottom).}
    \label{fig:compressiblity}
\end{figure}

Figure \ref{fig:compressiblity} shows the sorted magnitudes of the fitted coefficients (magnitude of the ECI times the multiplicity of the correlation function) for each expansion type and materials system for both the sparsest models and most accurate models obtained. Based on the adequacy of the resulting fit error metrics and the fast decay of coefficients shown in Figure \ref{fig:compressiblity}, we can conclude that the configuration energy if not exactly sparse is highly compressible---since signals or functions with power law decaying (or faster) coefficients can be well approximated by a small subset of terms.\cite{davenportIntroductionCompressedSensing2012} Specifically, a series of coefficients obey a power law decay if the sorted sequence satisfies the following,
\begin{equation}
    |c_i| \le Ci^{-q}
\end{equation}
where $c_i$ are the coefficients and $C, q > 0$ are constants. For all expansion types we see that the coefficient magnitude decay is faster than the power law decay shown.

We also see that in the results in Figure \ref{fig:compressiblity}, the coefficient decay of fits using the Potts frame are, at worst, the second fastest decaying series for both the sparsest and most accurate fits, such that expansions using the Potts frame are arguably more reliable in yielding sparser models than standard cluster expansions. This seems surprising considering that the total number of terms in the underdetermined system is exceedingly larger than that of the standard cluster expansions. However, considering the geometry of the union of $s$-dimensional subspaces as illustrated in Figure \ref{fig:subspaces}, we posit that indeed functions of configurational energy lie close to one of these subspaces with high probability, and we can therefore accurately represent the function with a much smaller number of terms than the total considered in the underdetermined system. Additionally, in light of the Theorem in Equation \ref{eq:cs-theorem}, the rapid decay of coefficients suggest that an $s$-sparse set of coefficients is close to the real coefficients and as a result the second term in the function approximation error is likely to be very small.

\begin{figure}
        \includegraphics[width=0.5\textwidth]{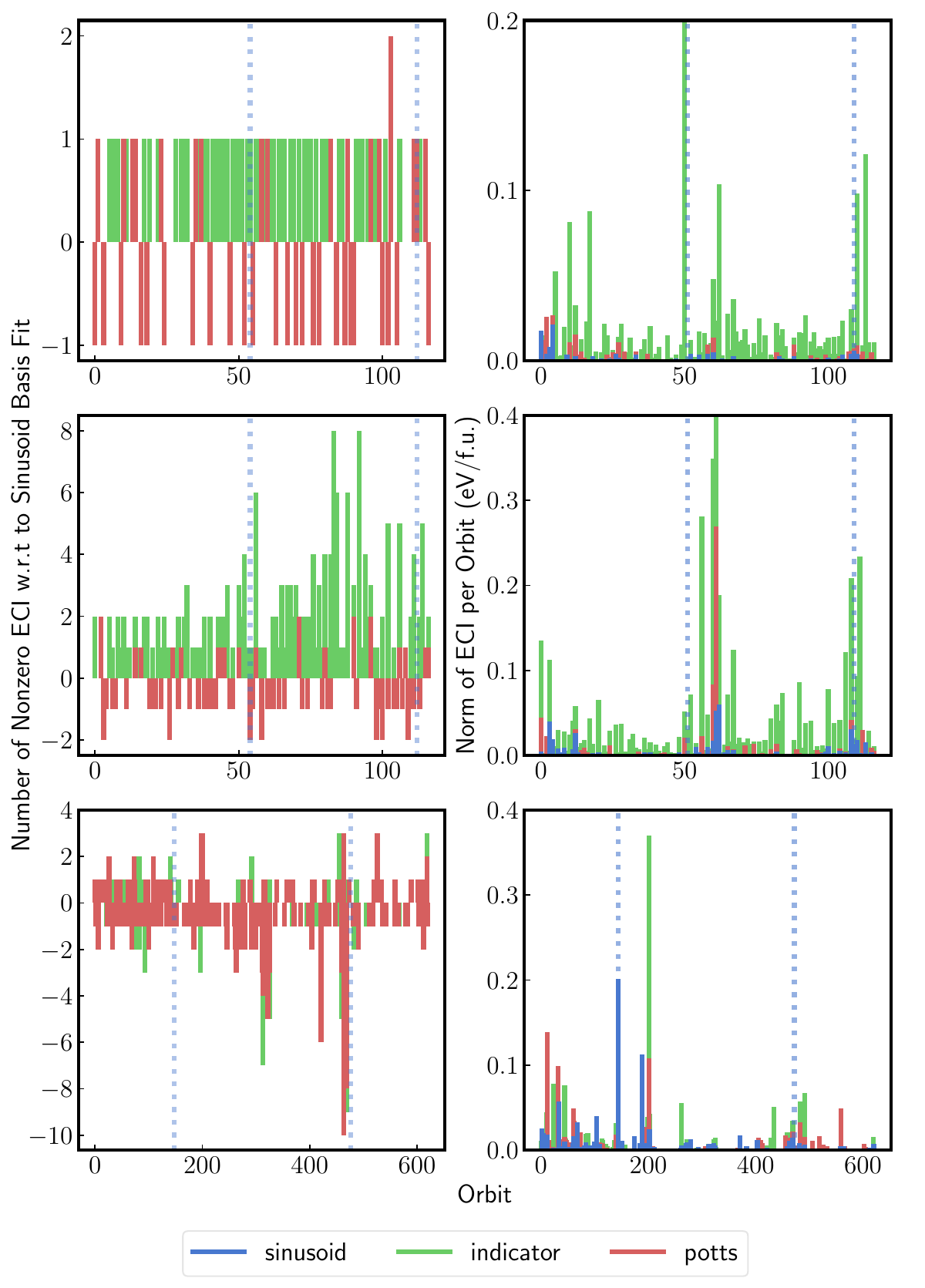}
    \caption{Number of nonzero coefficients relative to the sinusoid basis fit for each orbit, and norm of coefficients for each orbit for the most accurate models (Full RMSE). The vertical dotted lines separate the degree of orbit (pairs/triplets/quadruplets). LiMnOF binary-binary (top), LiMnTiOF ternary-binary (middle), LiMnOF quinary-ternary-binary (bottom).}
    \label{fig:spectra}
\end{figure}

We can take the previous results as ex post facto evidence that configuration energy is a compressible function and that CS with redundancy and coherent measurements works well for fitting expansions of configuration energy. Additionally, we observed that expansions fitted using the Potts frame also tend to follow expectations driven by physical considerations. Figure \ref{fig:spectra} shows the number of nonzero coefficients for each crystallographic orbit considered for the most accurate indicator basis and Potts frame based models with respect to the number of nonzero coefficients obtained in the most accurate sinusoid basis based model. We notice from the plots, that using the Potts frame, despite having a much larger number of total coefficients associated with each orbit, results in fits that set a similar number of nonzero coefficients within each orbit and never exceeds three additional coefficients per orbit when compared with the sinusoid CEM fit. The significance is that, not only do we recover accurate and sparse models, but the models themselves also have similar sparsity structure to the CEM based models. Figure \ref{fig:spectra} also shows the root sum of squares or norm of the fitted ECI for each orbit. We observed that for the most accurate models the fitted coefficient weight associated with each orbit is less erratic than that of the indicator based cluster expansion. Additionally, the fitted coefficients follow the well-known CEM heuristic of coefficient decay with orbit size more fittingly than the indicator based CEM. The aforementioned observations again demonstrate using the Potts frame can result in fitted expansions that are not only accurate and sparse, but also produce \textit{well-behaved} coefficients that align with practices and heuristics based on physical insights used in the field.

\section{Discussion \& Conclusion} \label{sec:conclusion}
We have briefly introduced the concepts behind CS with redundant representations and coherent measurements within the context of the cluster expansion method, with the goal of addressing the obstacles preventing adequate sampling of training structures to satisfy requirements of classical CS when fitting CE models. We proposed the generalized Potts frame as a redundant function set that can be used in lieu of standard CEM basis sets to represent functions over configurations of multicomponent crystalline materials. Additionally, we described the connections between the Potts frame and the commonly used nonorthogonal indicator function CEM basis set, as well as the essence of the Potts frame as a natural extension of the well known Potts model. Furthermore, we demonstrated the capability of obtaining arguably improved fitted expansions in terms of prediction accuracy, sparsity and robustness of coefficients compared to standard CEM fits. We stopped short of providing specific sensing and structure selection methods along with rigorous proofs that ensure small D-RIP constants; however the results from applications to complex lithium-transition metal oxide systems show promising results and the possibility of obtaining high accuracy expansions with such redundant representations.

Although this exposition of the power of using overcomplete frames in place of basis sets is particular in implementation and application, the concept can well be extended to many more problems in the growing space of novel methods for learning property-structure relationships in materials science. To our knowledge the concept has previously been used in the context of the CEM, although not explicitly, in the development of the variational basis cluster expansion and its connection to wavelets\cite{sanchez_cluster_2010}---which can formally constitute a redundant frame. As a matter of fact, the process of fitting a cluster expansion for any heterovalent ionic material system using only a charge-neutral configurations is itself an application of fitting an expansion using a frame. The reason being that a set of CEM correlation functions used in the space of functions over constrained configurations in heterovalent systems is in itself an overcomplete or redundant representation, thus the set of CEM correlation functions is rigorously a frame for the function space over such constrained configurations. Additionally, the application of redundant frames along with CS using coherent measurements may also be worthwhile in other areas of computational materials science which could benefit from higher representation flexibility and additional robustness, particularly in methods where statistical learning methods are used to estimate basis expansions in high dimensional spaces. The simple idea of using redundant sets to construct redundant product functions can be directly applied in cluster expansions beyond discrete species to include continuous vector spaces at each site such as in the spin and atomic cluster expansion.\cite{drautz_spin-cluster_2004, drautzAtomicClusterExpansion2019a} Particularly in the representation and learning of interatomic potentials, function basis sets over continuous vector spaces can be replaced with corresponding frames over those spaces. Furthermore, when configurational degrees of freedom exist, frames over continuous vector spaces can be combined with finite frames, such as a generalized Potts frame; and so in such cases the resulting product basis of an atomic cluster expansion becomes a tensor product of continuous and finite frames.

With the advent of the high entropy paradigm, as more complex materials systems are explored it will become increasingly necessary to develop methodologies that can work aptly with the rapidly growing multidimensional spaces and relatively fixed training data set sizes. The compressed sensing and sparse representation framework has been developed particularly for these situations in other fields and has previously been shown to be successful in simple materials science applications.\cite{nelsonCompressiveSensingParadigm2013, nelsonClusterExpansionMade2013} Continued exploration of systems with complexity beyond these systems requires continued development of methods to handle the increase in dimensionality. The generalized Potts frame developed and presented here represents one effort in said direction.

\acknowledgements
This work was primarily funded by the U.S. Department of Energy, Office of Science, Office of Basic Energy Sciences, Materials Sciences and Engineering Division under Contract No. DE-AC02-05-CH11231 (Materials Project program KC23MP). This material is also based upon work supported by the National Science Foundation Graduate Research Fellowship under Grant No. DGE 1752814.

A portion of the research was performed using computational resources sponsored by the Department of Energy's Office of Energy Efficiency and Renewable Energy and located at the National Renewable Energy Laboratory.

\appendix
\section{Frame Bounds for the Generalized Potts Frame} \label{sec:bounds}

\begin{definition}
A countable sequence $\{\Phi_\gamma\}_{\gamma\in I}$ is said to be a \textbf{frame} for a Hilbert space $\mathcal{H}$ if there exist frame bounds $A, B > 0$ such that, \cite{christensen_frames_2008, waldron_introduction_2018}
\begin{equation*}
    A||H||^2 \le \sum_{\gamma\in I}|\langle H, \Phi_\gamma\rangle|^2 \le B||H||^2\;\;\;\;\forall H\in\mathcal{H}
\end{equation*}
\end{definition}
The definition above implies that the frame $\{\Phi_\gamma\}_{\gamma\in I}$ spans $\mathcal{H}$.\cite{christensen_frames_2008}

To derive frame bounds for the generalized Potts frame, we work with the product-basis of cluster indicator functions without taking symmetry adapted averages. The bounds obtained apply equally to the functions obtained from symmetry adapted averages. The lower frame bound $A$ is obtained by splitting up the sum of projections into the frame elements as follows,
\begin{align*}
     \sum_{\gamma\in I}|\langle H, \bm{1}_\gamma\rangle|^2 &= \sum_{\gamma_{\max}}|\langle H, \bm{1}_{\gamma_{\max}}\rangle|^2 + \sum_{\gamma \in I\setminus\{\gamma_{\max}\}}|\langle H, \bm{1}_{\gamma}\rangle|^2 \\
     &= ||H||^2 + \sum_{\gamma \in I\setminus\{\gamma_{\max}\}}|\langle H, \bm{1}_{\gamma}\rangle|^2 \\
     &\ge ||H||^2
\end{align*}
Where $\{\gamma_{\max}\}$ denotes the set of all maximal clusters, and thus represents the canonical orthonormal basis for the Hilbert space $\mathcal{H}$. The second sum over all smaller clusters is always greater than or equal to zero, and so the obtained lower frame bound is $A = 1$. 

To obtain an upper frame bound we start by writing the Fourier expansion for cluster indicator functions,
\begin{align*}
    \bm{1}_\gamma(\bm{\sigma}) &= \sum_{\alpha}\langle\bm{1}_\gamma, \Phi_\alpha\rangle \Phi_\alpha(\bm{\sigma}) \\
    &= \sum_{\alpha}\prod_{i}\langle\bm{1}_{\gamma_i}, \phi_{\alpha_i}\rangle \Phi_\alpha(\bm{\sigma}) \\
    &= \sum_{\substack{\alpha;\\\text{supp}(\alpha)\subseteq\text{supp}(\gamma)}} \frac{1}{n^{\#\gamma}}\prod_{i}\phi_{\alpha_i}(\sigma_{\gamma_i})\Phi_\alpha(\bm{\sigma}) \\
    &= \sum_{\substack{\alpha;\\\text{supp}(\alpha)\subseteq\text{supp}(\gamma)}}\frac{1}{n^{\#\gamma}}\Phi_\alpha(\bm{\sigma_\gamma})\Phi_\alpha(\bm{\sigma}) \\
\end{align*}

where $\text{supp}(\cdot)$ represents the support or indices of the nonzero entries of $\alpha$ or $\gamma$, and $\#\gamma$ represents the total number of nonzero entries---i.e. the number of sites in a cluster. $n$ is the number of species allowed at a site and $\bm{\sigma}_\gamma$ is any occupancy string that includes the cluster represented by $\gamma$. Note the above expression is for a system with a single type of site (i.e. with the same set of allowed species in all sites). The general expression simply involves one over the product of the different number of species for each site instead of the factor $1/n^{\#\gamma}$.

Using the expansion given above for cluster indicator functions, we obtain an upper frame bound as follows,
\begin{align*}
    \sum_{\gamma\in I}|\langle H, \bm{1}_\gamma\rangle|^2 &= \sum_{\gamma\in I}\left|\left\langle H, \sum_{\substack{\alpha;\\\text{supp}(\alpha)\subseteq\text{supp}(\gamma)}}\frac{1}{n^{\#\gamma}}\Phi_\alpha(\bm{\sigma_\gamma})\Phi_\alpha(\bm{\sigma})\right\rangle\right|^2 \\
    &= \sum_{\gamma\in I}\left(\frac{1}{n^{\#\gamma}}\right)^2\left|\sum_{\substack{\alpha;\\\text{supp}(\alpha)\subseteq\text{supp}(\gamma)}}\hat{H}_\alpha\Phi_\alpha(\bm{\sigma_\gamma})\right|^2 \\
    &= \sum_{S}\frac{1}{n^{2|S|}}\sum_{\substack{\bm{\sigma}_\gamma;\\\text{supp}(\gamma)= S}}\left|\sum_{\substack{\alpha;\\\text{supp}(\alpha)\subseteq 
    S}}\hat{H}_\alpha\Phi_\alpha(\bm{\sigma_\gamma})\right|^2 \\
    &= \sum_{S}\frac{1}{n^{|S|}}\sum_{\substack{\alpha;\\\text{supp}(\alpha)\subseteq 
    S}}\hat{H}^2_\alpha \\
    &= \sum_\alpha \left(\sum_{S\supseteq\text{supp}(\alpha)}\frac{1}{n^{|S|}}\right)\hat{H}_\alpha^2 \\
    &\le\left(\sum_{S}\frac{1}{n^{|S|}}\right)||H||^2 \\
    &=\left(1+n^{-1}\right)^N||H||^2
\end{align*}
where $\hat{H}_\alpha$ are the Fourier coefficients of $H$. The sets $S$ contain site indices and the sum over these contains all possible subsets of site indices---i.e. all clusters of un-labeled sites. The upper frame bound obtained is $B = \left(1+n^{-1}\right)^N$, where $N$ is the total number of sites in the structure. The bound obtained is an improvement over the bound $2^N$ given by the Cauchy-Schwarz inequality.

\bibliography{refs.bib}

\end{document}